\documentclass[%
 aip,
 jmp,
 amsmath,amssymb,
 reprint,%
]{revtex4-1}
\usepackage{mathrsfs}
\usepackage{graphicx}
\usepackage{dcolumn}
\usepackage{bm}
\usepackage{caption}
\usepackage{CJK}
\usepackage{etoolbox}
\usepackage{float}
\usepackage{amstext}
\makeatletter
\def\@email#1#2{%
 \endgroup
 \patchcmd{\titleblock@produce}
  {\frontmatter@RRAPformat}
  {\frontmatter@RRAPformat{\produce@RRAP{*#1\href{mailto:#2}{#2}}}\frontmatter@RRAPformat}
  {}{}
}%
\makeatother
\begin{document}


\title[]{Giant Acceleration of Diffusion in Soft Matter Potential}
\author{Yu Lu
\begin{CJK}{UTF8}{gbsn}
(陆钰)
\end{CJK}
}
\affiliation{ 
School of Mechanical Engineering,
Nantong University,
Nantong 226002, China
}
\affiliation{ 
Shanghai Institute of Applied Mathematics and Mechanics, School of Mechanics and Engineering Science, Shanghai Key Laboratory of Mechanics in Energy Engineering,
Shanghai Frontier Science Center of Mechanoinformatics, Shanghai University,
Shanghai 200072, China
}
\author{Guo-Hui Hu
\begin{CJK}{UTF8}{gbsn}
(胡国辉)
\end{CJK}
$^*$}%
\email{ghhu@staff.shu.edu.cn}
\affiliation{ 
Shanghai Institute of Applied Mathematics and Mechanics, School of Mechanics and Engineering Science, Shanghai Key Laboratory of Mechanics in Energy Engineering,
Shanghai Frontier Science Center of Mechanoinformatics, Shanghai University,
Shanghai 200072, China
}
\affiliation{ 
Shanghai Institute of Aircraft Mechanics and Control, 100 Zhangwu Road, Shanghai 200092, China
}

\date{May 12, 2023}

\begin{abstract}
Diffusion of Brownian particles in the tilted periodic potential, usually referred to the washboard potential (WBP), is a well-known model to describe physical systems out of equilibrium.
Considering that the biological medium is flexible and thermally fluctuating, a new model, namely the soft matter potential (SMP), is proposed to describe the biological medium.
Compared to the washboard potential (WBP), SMP allows Brownian particles to actively modify the structure of the biological medium.
Brenner's homogenization theory is applied to predict the diffusivity and velocity of Brownian particles driven by external forces in SMP.
Thermodynamic uncertainty relation (TUR) is analyzed for Brownian particles in SMP.
It is found that, compared to WBP, Brownian particles in SMP require a lower energy cost $\langle q \rangle$ to achieve accuracy $\mathcal{A}$, i.e. Brownian particles in SMP have higher transport efficiency when driven by external forces.
\end{abstract}

\maketitle
\section{introduction}
The transport of Brownian particles in porous biological media is a topic of great relevance for life science and bio-medical engineering\cite{Natalie2022Massively,Kim2022Permeability,wang2013bursts,Hamann2019Nucleic,Ellis2001Macromolecular,Nakano2014EffectsOM,Tan2013MolecularCS,Nakano2017Model,Si2022Modulation,Si2018DNA}. Depending on the system, different transport properties of Brownian particles are desired. For example, in drug delivery systems, nanoparticles act as nanocarriers of drugs and need to reach the tumor site quickly and accurately. Once the nanocarrier approaches the lesion, the therapeutic molecule loaded on the nanoparticle with porous structure, such as gel or metal-organic framework, needs to be released rapidly\cite{liu2020Jaggregate,Zelepukin2022FlashDR,Culebras2021Wood}. On the other hand, for sustained-release systems\cite{Freitag2002randomized_extendedrelease,KANE2007Treatment}, the drug efficacy depends on its stable and slow release. Therefore, a deeper theoretical understanding of the transport of Brownian particles in complex environments is essential to solve various biomedical engineering problems.

The motion of Brownian particles is constrained by the local structure of the biological medium, and local constraints can be considered as a periodic potential $U$ \cite{Dell2013Theory,Xue2020Diffusion,cai2015hopping,Lu2021potential}, which greatly affects the drug utilization.
To improve the therapeutic effect, a particularly simple way is to manipulate Brownian particles by external forces $F_e$.
In the presence of a constant external force, Brownian particles can be equivalent to diffusion in a washboard potential, which is an widely-used model for non-equilibrium statistical physics\cite{Kramers1940Brownian,Gang1996Diffusion,Reimann2002Diffusion,Lindenberg2005Transport,Lindenberg2007Dispersionless,Reimann2008Weak,Burada2009Diffusion,Xiao2019Investigation,Reimann2002Diffusion,Lindenberg2005Transport}.

Over the last two decades, extensive investigations have been conducted to explore the diffusion of particles in a washboard potential\cite{hanggi1990reaction,Reimann2001Giant,Reimann2002Diffusion,Lindenberg2005Transport,Lindenberg2007Dispersionless,Reimann2008Weak,Burada2009Diffusion,Bellando2022Giant}.
Based on an overdamped nonlinear Langevin equation, Reimann {\it et al.}\cite{Reimann2001Giant} studied the Brownian motion of particles in a static tilted periodic potential or washboard potential (WBP), and used the first passage time theory\cite{hanggi1990reaction} to derive accurate expressions for the average velocity and diffusivity of particles.
Their theoretical results indicated that the diffusivity as a function of the tilted force exhibits a peak larger than the free diffusivity $D_0$ near the critical force, which is called the giant acceleration of diffusion (GAD).
Considering that the tiny and time-independent deviations of the periodic potential are unavoidable in experimental realizations, Reimann {\it et al.}\cite{Reimann2008Weak} studied the GAD of particles diffuse in the WBP with weak disorder.
Their theoretical results implied that even a tiny amount of disorder in WBP may further boost the particle diffusion when a tilted force near the critical force.
Burada {\it et al.}\cite{Burada2009Diffusion} studied the diffusion of particles in a channel with periodically varying cross sections, which can be considered as a periodic entropy potential.
Based on the theoretical study by Reimann {\it et al.}\cite{Reimann2001Giant,Reimann2008Weak}, they obtained the theoretical results of the velocity and diffusivity of the particles.
Based on the Fokker-Planck equation, Kim {\it et al.} \cite{Kim2022Permeability} studied the solvent permeation through a polymer membrane under external force driving using numerical simulation. They simplified the effect of the polymer membrane on the solvent as a potential filed, and obtained the steady-state density distribution of the solvent that depends on the external force.

Driven by the external force, the particle diffusion in periodic potential can be viewed as the physical systems deviate from equilibrium state.
Barato {\it et al.}\cite{Barato2015Thermodynamic} proposed the thermodynamic uncertainty relation (TUR), which is a quantitative tool to characterize systems out of equilibrium\cite{horowitz2020thermodynamic,Koyuk2020Thermodynamic}.TUR is an inequalities that give the limits on the precision of particle transport for a certain energy dissipation.
Hyeon {\it et al.}\cite{Hyeon2017Physical} used TUR to analyze the Brownian motion in periodic potentials $U$ tilted by a force $F_e$.
The average travel distance $\langle x_n(t) \rangle$ of particles has its variance $\delta\langle x_n^2(t) \rangle=\langle x_n^2(t) \rangle-\langle x_n(t) \rangle^2$, which can be used to define the transport accuracy $\mathcal{A}=\langle x_n(t) \rangle^2/\delta\langle x_n(t)^2 \rangle$.
The work done by the external force is the heat dissipation $\langle q \rangle = F_{e} \langle x_n(t) \rangle$.
The ratio of heat dissipation and transport precision follows the thermodynamic uncertainty relation (TUR), that is, $\mathcal{Q}=\langle q \rangle/\mathcal{A} \geq 2k_B\Theta$.
For an aimed transport precision, the heat dissipation decreases with the decrease of $\mathcal{Q}$, but there is also a lower bound $2\mathcal{A}k_B\Theta$.
As a quantitative tool, $\mathcal{Q}$ is used to describe transport efficiency.

Compared with the solid surfaces\cite{Frenken1985Observation}, optical fields\cite{Lee2006Giant} and confined geometries\cite{Burada2009Diffusion}, the biological medium is non-static and has two distinct differences:
(1) The thermal fluctuation of biological medium such as mucus, cyto-skeletons and brain-blood barriers will affects the transport of particles embedded in such biological medium;
(2) The biological medium is flexible, the particle can actively change the structure of the medium. Therefore, a static washboard potential (WBP) that only varies with the coordinates of the particles $x_n$ hardly accurately describe the biological medium, especially the interaction between particles and biological medium.

A typical example of such a biological medium can be modeled by a polymer network\cite{Dell2013Theory,Xue2020Diffusion,cai2015hopping,Lu2021potential,Xu2021Enhanced}.
In our recent study\cite{Lu2021potential,Lu2022Double}, particle diffusion was investigated in an unentangled polymer network with periodic structure.
It is found that the junctions in the polymer network fluctuated around their origin locations.
When particles try to hop in the polymer network, the particles actively modify the structure of network, which is considered as  junctions deviation.
Therefore, in addition to the loop stretching proposed by Cai {\it et al.}\cite{cai2015hopping}, junction deviation is a crucial mechanism for particle diffusion in the polymer network.
Based on these results, a simplified double spring model\cite{Lu2022Double} was proposed to calculate potential $U$ for particles diffusion in polymer network.
In the double spring model, the particles $x_n$ and the junction $x_c$ are connected by a linear spring, and the junction $x_c$ is constrained by a polymer network, which is described by a non-linear spring.
The potential $U$ is a function of the coordinates of the particles and junctions $x_n$ and $x_c$.
This idea can be extended to the general biological medium, the variable $x_c$, namely medium coordinates, is a reference mass point in the biological medium.
To explore the diffusion of particles in biological medium under external force, a tiled periodic soft matter potential (SMP) is proposed in the present investigation, in which the medium coordinates is introduced to consider the coupling effect between the particles and the medium, i.e. particle can actively modify the structure of medium.
Specifically, the coordinates of the medium $x_c$ characterize the thermal fluctuation and flexibility of the biological medium.

In the present work, the double spring model is utilized to construct a soft matter potential (SMP) to describe particle diffusion in a biological medium.
 Brownian particles diffusion in SMP with tilted force $F_e$ is studied by overdamped nonlinear Langevin equations, in comparison to the results of WBP.
The Langevin equations are solved by a Brownian dynamics simulation to obtain the temporal evolution of the mean displacement of the particles $\langle x_n(t) \rangle$ and the variance of displacement $\delta\langle x_n^2(t) \rangle$.
Brenner's homogenization theory is applied to predict the long-time transport properties (diffusivity $D_L$ and velocity $v_L$) of Brownian particle diffusion in WBP and SMP.
Finally, the TUR of particle diffusion in WBP and SMP is analyzed and compared.

\section{Method and Model}
\subsection{Washboard Potential}
Considering a particle $x_n$ driven by a stochastic force moving in a periodic potential $V(x_n)$ of amplitude $V_0$ and period $L$, and subject to constant external force $F_e$.
The coordinate of Brownian particle $x_n$ is governed by an over-damped non-linear Langevin equations:
\begin{equation}\label{equ:ONLE_WBP}
\left\{
\begin{aligned}
    &dx_n+D_0\frac{\partial V-F_ex_n}{\partial x_n}dt=\sqrt{2D_0}d\mathcal{W}_n; \\
    &V(x_n)=V_0(2x_n/L)^2, x_n\in [-L/2,L/2];\\
    &V(x_n+L)=V(x_n).
\end{aligned}
\right.
\end{equation}
in which $\mathcal{W}_c$ is the wiener process, $D_0$ is the diffusivity for particle diffusion in simple fluid.
The potential $V$ represents the interaction between particle and the static medium, which is a function of the particle coordinates $x_n$.
The tilted periodic potential $V(x_n)-F_ex_n$ is called washboard potential.

\subsection{Soft Matter Potential}
In the present study, we present a physical model which is capable of describing thermal fluctuation and deformation of the biological medium.
One idea is to introduce non-constant medium coordinates $x_c$.
For this purpose, the nonstatic interaction potential $V_s(x_n,x_c)=V_0(2d_{nc}/L)^2$ is defined as a function of distance $d_{nc}=|x_c-x_n|$, to model the loop stretching.
When $x_c = 0$, the medium is static, the interaction returns to a static potential $V_s(x_n,0)=V(x_n)$.

In our previous study\cite{Lu2022Double}, the medium coordinates $x_c$ is used to describe the junctions deviation during particle diffusing in polymer network.
The junctions in polymer network is constrained by polymer chains, which is described by a confinement potential:
\begin{equation}\label{equ:VC}
W(x_{c})=W_0(2x_c/L)^{\alpha},
\end{equation}
where $W_0$ represent the medium rigidity.
As the rigidity $W_0$ increase, the biological medium becomes harder to resist on the deformation.
$\alpha=1.5$ is the nonlinear factor based on our previous study\cite{Lu2022Double}.
Without loss of generality, the confinement potential $W$ is used to characterize the constraints of the biological medium on the medium coordinates $x_c$.
The total potential $U=V_s+W$ in our model is called the soft matter potential (SMP).

In the soft matter potential, the coordinate of Brownian particles $x_n$ and the medium coordinate $x_c$ is governed by two non-linear Langevin equations:
\begin{equation}\label{equ:ONLE_SMP}
\left\{
\begin{aligned}
    &dx_n+D_0\frac{\partial U}{\partial x_n}dt-D_0F_edt=\sqrt{2D_0}d\mathcal{W}_n; \\
    &dx_c+D_c\frac{\partial U}{\partial x_c}dt=\sqrt{2D_c}d\mathcal{W}_c,
\end{aligned}
\right.
\end{equation}
in which $\mathcal{W}_c$ is a Wiener process for the medium coordinate.
$D_c$ is the free diffusivity of the reference mass point in biological medium $x_c$, which represents the thermal fluctuation of the biological medium.
The corresponding Fokker-Planck function of equation (\ref{equ:ONLE_SMP}) is given by
\begin{equation}\label{equ:FP_SoftPEL}
\left\{\begin{aligned}
&\frac{\partial \rho(x_n,x_c,t)}{\partial t}=-\frac{\partial  J_n}{\partial x_n}-\frac{\partial  J_c}{\partial x_c}= \mathscr{L}_{s} \rho(x_n,x_c,t),  \\
&J_n=D_0 \left[ -\frac{\partial U}{\partial x_n} +F_e  -\frac{\partial }{\partial x_n}  \right]\rho(x_n,x_c,t), \\
&J_c=D_c \left[ -\frac{\partial U}{\partial x_c}   -\frac{\partial }{\partial x_c}  \right]\rho(x_n,x_c,t), \\
\end{aligned}\right.
\end{equation}
where $\rho$ is the joint probability density distribution function (PDF) of displacement $x_n$ and the medium coordinate $x_c$, $\mathscr{L}_{s}$ is the Fokker-Planck operator of the soft matter potential. $J_n$ and $J_c$ are the probability currents.

\subsection{Brownian Dynamics}
Brownian dynamics simulations can be applied to solve equations (\ref{equ:ONLE_WBP}) and (\ref{equ:ONLE_SMP}). The simulation domain is one periodic $L$ with specific potential landscape. By using the periodic boundary condition, the particle coordinate and medium coordinate is 
\begin{equation}
    x_n=\lambda_n L+ \xi_n,\ \  x_c=\lambda_c L+ \xi_c.
\end{equation}
in which 
$\xi_n$ and $\xi_c$ is the position of particle and reference mass point of biological medium in simulation domain, $\lambda_n,\lambda_c=...,-2,-1,0,1,2,...$ represents the mirror computational domains. 
Ensemble averaging is performed on the system to obtain the average displacement $\langle x(t)\rangle$ and the displacement variance $\delta \langle x^2(t)\rangle$ with time $t$.  All cases are calculated with ensembles $2\times 10^5$ under different initial conditions.

\subsection{Prediction of long time stage properties}
In long time stage, particle diffusion in the WBP will migrate with the velocity\cite{Reimann2001Giant},
\begin{equation}\label{equ:vL}
\begin{aligned}
v_L=\lim_{\infty} v(t)=\lim_{\infty} \langle x_n(t) \rangle/t,
\end{aligned}
\end{equation}
and diffusivity $D_L$ is given by,
\begin{equation}\label{equ:DL}
D_L=\lim_{t\rightarrow \infty}D(t)=\lim_{t\rightarrow \infty}\frac{\delta\langle x^2(t) \rangle}{2t},
\end{equation}
in which displacement variance $\delta\langle x^2(t) \rangle=\langle x^2(t) \rangle-\langle x(t)\rangle^2$.

Adrover {\it et al.}\cite{Adrover2019Exact,Adrover2019Laminar} studied the laminar dispersion in a sinusoidal microtube.
As a convection-diffusion transport problem\cite{wang2022double,Jakhar2023Instability,Vargas2023ColloidPOF,Venditti2022Exactdispersion,Venditti2022ComparisonPOF}, Brenner's homogenization theory is applied to calculate the long-time velocity and diffusivity.
By using their homogenization method, Venditti {\it et al.}\cite{Venditti2022Physica} studied the motion of Brownian particles in WBP in inertial regime.
In the present work, the Brenner's homogenization theory is utilized to calculate $v_L$ and $D_L$ of Brownian particles in SMP.

Equation (\ref{equ:FP_SoftPEL}) can be simplified to a constant coefficient convection-diffusion equation in the long time scale:
\begin{equation}\label{equ:FP_EDC_xn}
\frac{\partial \rho_n(x_n,t)}{\partial t}=\frac{\partial }{\partial x_n}\left[-v_{L}+D_{L}\frac{\partial }{\partial x_n}\right]\rho_n(x_n,t),
\end{equation}
in which $\rho_n(x_n,t)$ is the PDF of particle coordinates
\begin{equation}
\rho_n(x_n,t)=\int_{-\infty}^{\infty}\rho(x_n,x_c,t)dx_c.
\end{equation}
Based on the idea of homogenization method, the average velocity $v_L$ and diffusivity $D_L$ can be calculated by \cite{Brenner1980Dispersion}:
\begin{equation}\label{equ:EDC_vldl}
\left\{
\begin{aligned}
 &v_L=\int_0^{L}\int_0^{L}(F_e-\frac{\partial U}{\partial x_n})w_0(\xi_n,\xi_c)d\xi_n d\xi_c, \\ 
&D_L=\int_0^{L}\int_0^{L}(F_e-\frac{\partial U}{\partial x_n}-v_L)w_0(\xi_n,\xi_c)b(\xi_n,\xi_c)d\xi_n d\xi_c,
\end{aligned}\right.
\end{equation}
where $w_0$ is the PDF of local coordinates $\xi_n$ and $\xi_c$, i.e., $w_0=\overline{\rho}(\xi_n,\xi_c)=\sum_{\lambda_n,\lambda_c}\rho(\xi_n+\lambda_nL,\xi_c+\lambda_cL)$, satisfying the homogeneous equation and normalization condition,
\begin{equation}\label{equ:EDC_w0}
\mathscr{L}_sw_0=0, \ \ \int_0^{L}\int_0^{L} w_0 d\xi_nd\xi_c=1.
\end{equation}
with periodic boundary conditions at $\xi_n=0,L$
\begin{equation}\label{equ:EDC_w0_BD}
w_0(0,\xi_c)=w_0(L,\xi_c), \ \ \frac{\partial w_0}{\partial \xi_n} \bigg|_{\xi_n =0} = \frac{\partial w_0}{\partial \xi_n} \bigg|_{\xi_n =L}.
\end{equation}
$b(\xi_n,\xi_c)$ is the so-called $b-filed$\cite{Brenner1980Dispersion}, representing the deviation of the position of the particle from the mean position $v_Lt$.
The $b-filed$ is an solution of the non-homogeneous function:
\begin{equation}\label{equ:EDC_wb}
\mathscr{L}_s w_0b=-(F_e-\frac{\partial U}{\partial \xi_n}-v_L)w_0 +2D_0 \frac{\partial w_0}{\partial \xi_n}.
\end{equation}
equipped with periodic boundary conditions at $\xi_n=0,L$.


\section{Results}

\subsection{Temporal evolution of velocity and diffusivity}
Brownian dynamics simulations is conducted to study the diffusion characteristics of particles in two types of potential (SMP and WBP) at different time scales for different tilted force $F_e$.
The tilted force $F_e$ is normalized by the critical force $F_0=2V_0/L$.
The dimensionless force $\tilde{F}=F_e/F_0$ is a quantitative measure of the equilibrium state of the system.
$\tilde{F}\ll 1$ indicates the system near the equilibrium, the tilted potential has wells and barriers, and the particle is confined to the potential from which a thermal fluctuation causes it to escape.
When $\tilde{F}= 1$, the maxima and minima of tilted potential disappear.
$\tilde{F}\gg 1$ means the system out of equilibrium, periodic portion of the potential becomes unimportant\cite{Lindenberg2007Dispersionless}.
Parameters are set to $D_0=1,V_0=6.25$ both for WBP and SMP in the present study.
Additional parameters for SMP are set to $D_c=D_0,W_0=V_0$.

\begin{figure}
	\centering
	\includegraphics[width=0.40\textwidth]{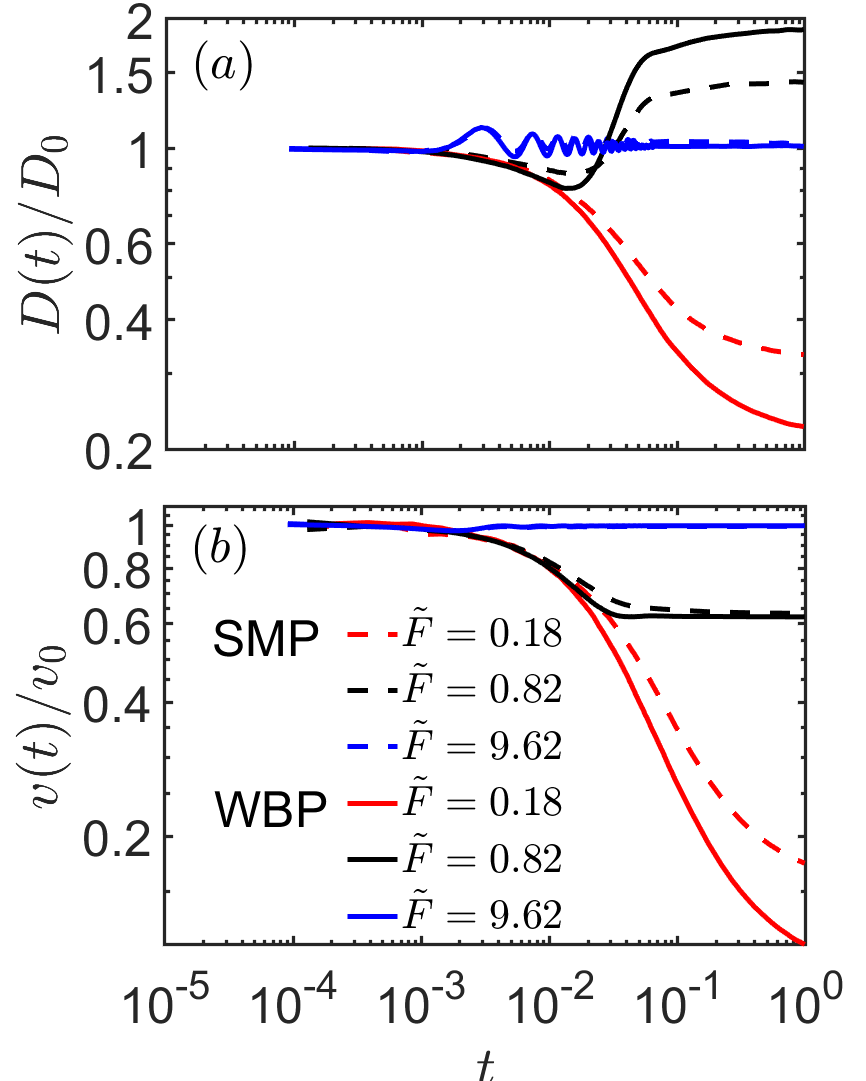}
 \captionsetup {justification=raggedright,singlelinecheck=false}
	\caption{Temporal evolution of normalized diffusivity $D(t)/D_0$ (a) and velocity $v(t)/v_0$ (b) of particle diffusion in WBP (solid line) and SMP (dash line).Parameters are set to $D_0=1,V_0=6.25$ both for WBP and SMP.The additional parameters for SMP are set to $D_c=D_0,W_0=V_0$.
}

	\label{img:DtVt_WBP_SMP}
\end{figure}

As shown in Fig. \ref{img:DtVt_WBP_SMP}, the temporal evolution of diffusivity $D(t)/D_0$ and velocity $v(t)/v_0$ exhibits three stages\cite{hofling2013anomalous,Lu2022Double,Lu2021potential},in which $D_0$ and $v_0=F_eD_0$ are the transport properties of the particle in a homogeneous medium.
In the short time regime, for both SMP and WBP, the diffusivity and the velocity are constant, equal to the values of the diffusivity and velocity of particles in a homogeneous medium. At this spatiotemporal stage, the complicated environment defined by the potential field has not yet affected the motion of the particles.

After the short time regime, the temporal evolution of diffusivity and the velocity exhibit dependence on the type of potential and the external force $F_e$.
When tilted force lower or near the critical force, the velocity $v(t)$ exhibits a transition from sub-transport to free transport long time velocity $v_L$.
Compared to the velocity $v_0$ of particles in simple fluid, the velocity $v_L$ is reduced for particles in titled potential.
As the external force $F_e$ increases, $v_L$ gradually approaches the $v_0$ when $\tilde{F}\gg
1$.
Besides, the temporal evolution of velocity is not affected by the type of potential except that $\tilde{F}\ll 1$.

When the tilt force is lower than the critical force, $D(t)$ exhibits a transition from subdiffusion to free diffusion with reduced diffusivity $D_L<D_0$.
While $D(t)$ exhibits a transition from superdiffusion to free diffusion with enhanced diffusivity $D_L>D_0$ when the external force is close to the critical force.
Similarly to $v(t)$, for a higher tilt force $\tilde{F}\gg 1$, $D(t)$ is constant as time elapses and $D_L=D_0$.
These results indicate that, as the tilted force increase, the long-time diffusivity does not change monotonically but has a peak when $\tilde{F}\approx 1$, which is so-called giant acceleration of diffusion (GAD).

When the external force is smaller than and close to the critical force, the type of potential has a significant effect on the diffusivity. Compared to WBP, the particles diffusing in the SMP show a much weaker sub-diffusion ($\tilde{F} \ll 1$) and super-diffusion ($\tilde{F}\approx 1$) at intermediate time scales. 
As a result, compared with the long-time diffusivity $D_L$ of particle in WBP, $D_L$ of the particles in the SMP is smaller when $\tilde{F}\approx 1$ and larger when $\tilde{F} \ll 1$.
The results indicate that the motion of particles in SMP is weakly dispersed, the GAD is weakened in a flexible and fluctuating medium.

\subsection{Giant acceleration of diffusion in SMP}
For particle diffusion in a WBP, giant acceleration of diffusion (GAD)  has been extensively investigated by previous experimental and theoretical studies, and confirmed by the results from Brownian dynamics\cite{Reimann2001Giant,Lee2006Giant,Reimann2008Weak,Burada2009Diffusion}.
By applying Brenner's homogenization theory, the long-time diffusivity $D_L$ and velocity $v_L$ for different $F_e/F_0$ is widely used to analyse the GAD phenomena.
In the present study, the GAD phenomena of particles diffusing in a SMP are explored with particles in a WBP with the same interaction $V_0$.
Parameters are set to $D_0=1,V_0=6.25$ both for WBP and SMP.
The additional parameters for SMP are set to $D_c=D_0,W_0=V_0$.

\begin{figure}[h]
	\centering
	\includegraphics[width=0.45\textwidth]{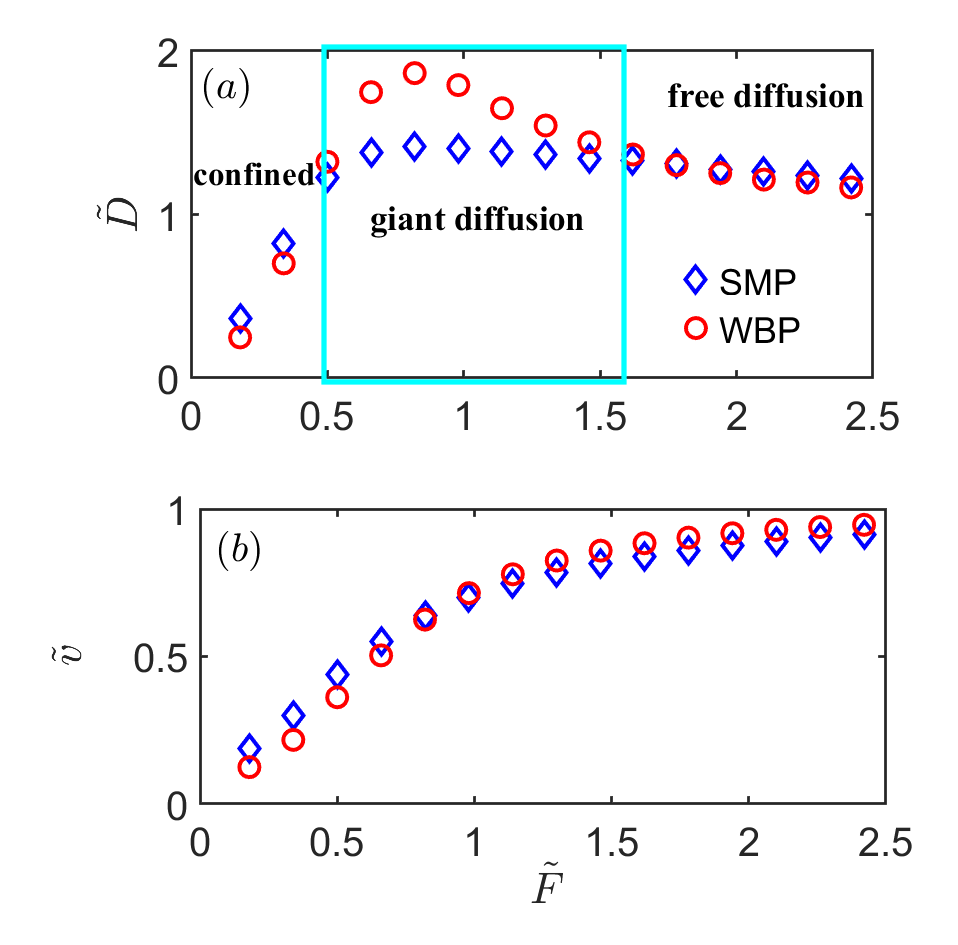}
  \captionsetup {justification=raggedright,singlelinecheck=false}
	\caption{
 The variety of long time normalized diffusivity $\tilde{D}=D_L/D_0$ (a) and velocity $\tilde{v}=v_L/v_0$ (b) with external force $F_e$ for particle diffusion in WBP (red cycle) and SMP (blue diamond) for the same parameters as in Figure \ref{img:DtVt_WBP_SMP}.
}
	\label{img:WBP_SMP_DL_vL}
\end{figure}

The variation of normalized diffusivity $\tilde{D}(\tilde{F})=D_L/D_0$ and velocity $\tilde{v}(\tilde{F})=v_L/v_0$ with $\tilde{F}$ is shown in Fig. \ref{img:WBP_SMP_DL_vL}.
It is found, for smaller external force, approximately $\tilde{F} < 0.5$, both particle diffusivity $\tilde D$ and its velocity $\tilde{v}$ are less than 1  for SMP and WBP, i.e., $\tilde{D}<1$, $\tilde{v}<1$, indicating that the particle is well confined in the potential.
The diffusivity and velocity of particles in SMP is slightly larger than those of WBP with the same $\tilde{F}$.
For intermediate external force $\tilde{F}$, i.e., the tilted force is close to the critical force $\tilde{F}\approx 1$, particles in both WBP and SMP exhibit maximum diffusivity, and the diffusivity peak for WBP is obviously larger than SMP, as shown in Fig. \ref{img:WBP_SMP_DL_vL}(a). However, Fig. \ref{img:WBP_SMP_DL_vL}(b) suggests that the variation of normalized velocity $\tilde{v}$ is similar for WBP and SMP, although there are distinct differences in terms of diffusivity.
When the system is far from equilibrium with the titled force significantly larger than the critical force ($\tilde{F}\gg 1$), the influence of the confinement from potential well on particle diffusion becomes negligible. The particle motion is akin to rolling down a slope and the long time transport properties approach the results in homogeneous medium, i.e. $D_L \approx D_0$ and $v_L \approx v_0$.

\subsection{Influence of the biological medium properties}
The particles diffusing in the SMP exhibit transport characteristics different from those that diffuse in the WBP, which originate from the deformability and thermal fluctuations of the biological medium characterized by the SMP.
In the SMP model, the deformability and thermal fluctuations of biological medium is quantitatively described by two parameters: (1) the thermal fluctuations strength of the biological medium, which is controlled by $D_c$; (2) the stiffness of the biological medium, which is the parameter $W_0$ in equation (\ref{equ:VC}).

\begin{figure*}[]
	\centering
	\includegraphics[width=0.8\textwidth]{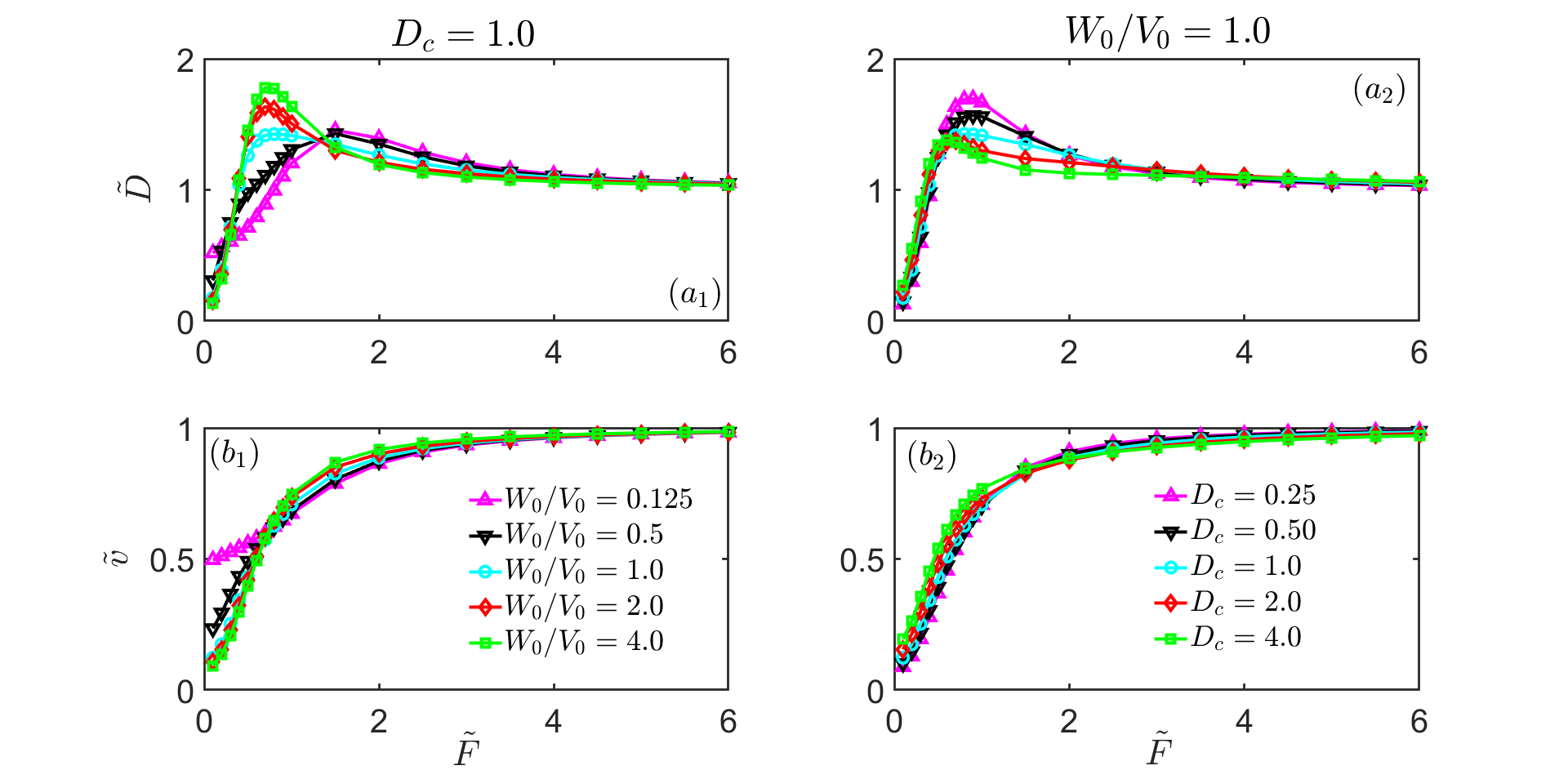}
  \captionsetup {justification=raggedright,singlelinecheck=false}
	\caption{
         The variation of the diffusivity ($a_1$) ($a_2$) and velocity ($b_1$) ($b_2$) of the particles with the external force in SMP when $D_c$ and $W_0$ is varied.
The SMP parameters set to $D_0=1$ and $V_0=6.25$, and $W_0$ (left) and $D_c$ (right) are varied.
	}
	\label{img:vldl_kcDc}
\end{figure*}

As shown in Fig. \ref{img:vldl_kcDc}, the variation of the velocity $\tilde{v}(\tilde{F})$ and diffusivity $\tilde{D}(\tilde{F})$ of the particles with the external force is calculated by the Brenner's homogenization theory.
The SMP parameters set to $D_0=1$ and $V_0=6.25$, and the dependence of the parameters on $D_c$ and $W_0$ are analyzed.

The influence of biological medium properties ($W_0$ and $D_c$) on normalized velocity $\tilde{v}$ is negligible expect that the stiffness is small ($W_0=0.125V_0$) at near-equilibrium ($\tilde{F}\ll 1$).
For normalized diffusivity $\tilde{D}$, when $W_0$ decreases and $D_c$ increases, $\tilde{D}$ increases near equilibrium ($\tilde{F}\ll 1$) and decreases near the critical force ($\tilde{F}\approx 1$).
When the system is far away from equilibrium ($\tilde{F}\gg 1$), $W_0$ and $D_c$ do not affect the transport properties of particles, diffusivity and velocity approach the results of particles in a simple fluid.

These results indicate that smaller stiffness $W_0$ and stronger thermal fluctuation $D_c$ make the biological medium approach a simple fluid, while the opposite makes the SMP closer to a WBP.
In addition, the properties of the biological medium have a profound influence on the diffusivity than on the velocity.

\subsection{Thermodynamic uncertainty relation of Brownian particle in SMP}
In the design of drug delivery, we expect to achieve rapid and precise transport of drug-loaded nanocarrier, while minimizing energy consumption in the process, that is, optimizing the transport efficiency.
For sustained-release systems, we expect to be able to flexibly control their release rate.

Thermodynamic uncertainty relation (TUR) proposed by Barato {\it et.al.}\cite{Barato2015Thermodynamic} is a quantitative tool are available for the description of transport efficiency.
The key idea of TUR is that the ratio of energy dissipation and accuracy has a lower limit $2k_BT$.
The energy dissipation of the system can be described by the work done by the external force $F_e$, that is, $\langle q \rangle = F_{e} \langle x_n(t) \rangle$. The precision $\mathcal{A}$ is given by
\begin{equation}
\mathcal{A} = \frac{\langle x_n(t) \rangle^2}{\langle x_n(t)^2 \rangle -\langle x_n(t) \rangle^2}.
\end{equation}

At long time stage, both the energy dissipation $\langle q \rangle$ and the accuracy $\mathcal{A}$ grow linearly with time.
The ratio between $\langle q \rangle$ and $\mathcal{A}$, noted as $\mathcal{Q}$, is a function of \textit{}tilted force and independent on the time, which is given by
\begin{equation}\label{equ:CPt_Qlt}
\mathcal{Q}(\tilde{F})/k_BT=\frac{2F_eD_L}{v_L}=2\frac{\tilde{D}(\tilde{F})}{\tilde{v}(\tilde{F})}\geq 2.
\end{equation}
As a quantitative tool, $\mathcal{Q}$ can be used to describe the efficiency of transport.
A lower $\mathcal{Q}$ implies that the expected transport accuracy can be achieved through a smaller energy dissipation.

\begin{figure}[h]
	\centering
	\includegraphics[width=0.45\textwidth]{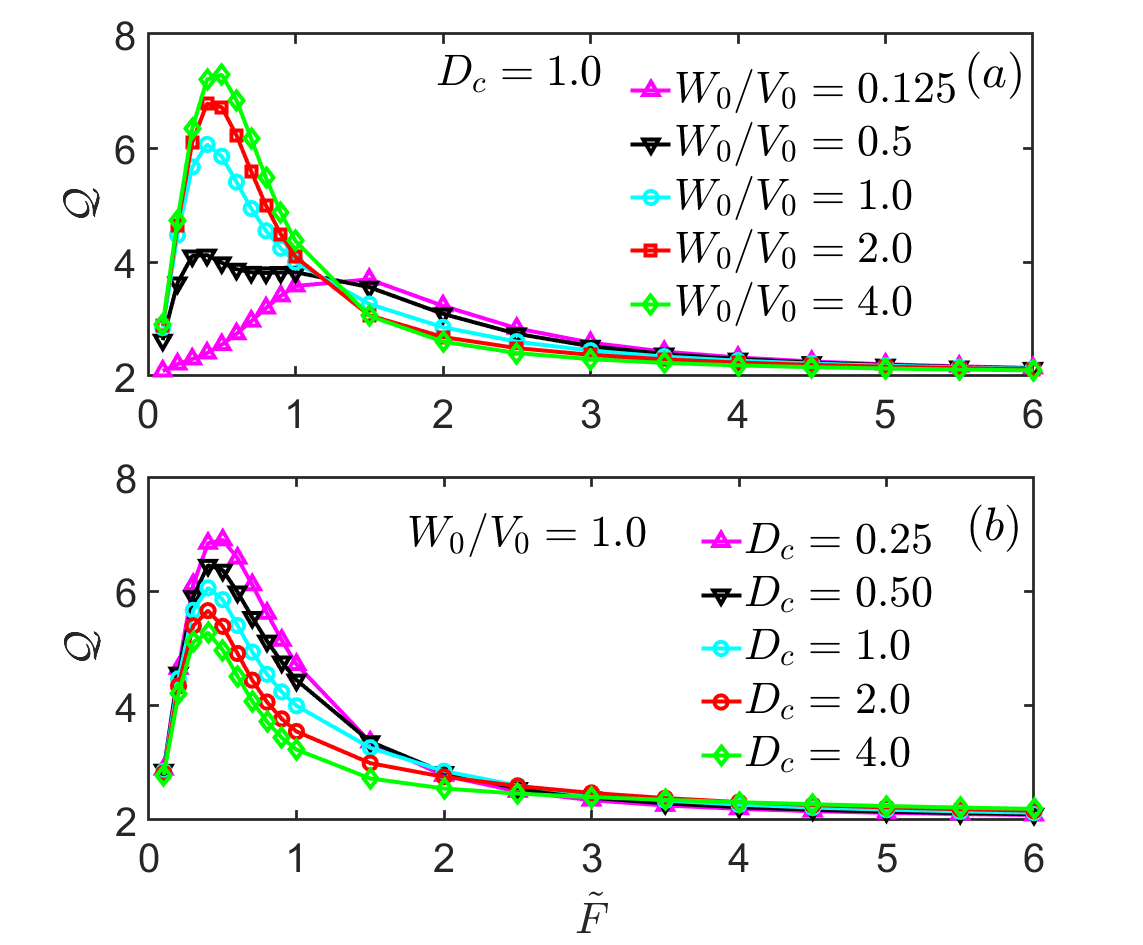}
  \captionsetup {justification=raggedright,singlelinecheck=false}
	\caption{
        Function $\mathcal{Q}$ with tilt $\tilde{F}$ for particle diffusion in SMP ($D_0=1$ and $V_0=6.25$).
        (a)For a fixed $D_c=1.0$, the variety of $\mathcal{Q}(\tilde{F})$ for different $W_0$.
        (b)For a fixed $W_0/V_0=1.0$, the variety of $\mathcal{Q}(\tilde{F})$ for different $D_c$.     
    }
	\label{img:TUR_kcdc.png}
\end{figure}

As shown in Fig. \ref{img:TUR_kcdc.png}, function $\mathcal{Q}$ exhibit a peak when $\tilde{F}\approx 1$ and reach the lower limit at weak titled $\tilde{F} \ll 1$ and strong titled $\tilde{F}\gg 1$\cite{Hyeon2017Physical}.
These results indicate that the transport efficiency is optimized with very large and negligible external forces.
However, a tiny external forces rapidly reduce the velocity $v_L$, low transport speed may decrease the therapeutic effect.
Very large external forces are difficult to apply or damage the biological medium, which means that the external force has an upper constraint.
Therefore, selecting an external force near the critical force is a reasonable way to optimize the transport efficiency.
However, the peak of $\mathcal{Q}$ indicates that the particle needs more energy cost $\langle q \rangle$ to achieve the target accuracy $\mathcal{A}$ when $\tilde{F}\approx 1$.

Particles moving in SMP with a smaller stiffness $W_0$ and stronger thermal fluctuations $D_c$ have lower diffusivity peaks and maintain the velocity unchanged, implying a smaller $\mathcal{Q}(\tilde{F}\approx 1))$ when the external force is near the critical force.
As shown in Fig. \ref{img:TUR_kcdc.png}, the peak of $\mathcal{Q}$ decreases when $W_0$ decreases and $D_c$ increases.
The flexible structure of the biological medium leads to a higher transport efficiency of the particle and ensures the transport speed.
This implies that, compared to particles in the WBP, particles in the SMP have higher transport efficiency. Moreover, the transport efficiency increases with the intensity of thermal motion and the decrease in stiffness in the SMP.
This result will provide important theoretical support for drug delivery.

\section{Conclusion}
In present work, the transport of particles in the biological medium driven by external force is studied utilizing the numerical and theoretical method. 
The washboard potential (tilted periodic potential) is a well-known model to study transport of particles in complex medium.
Taking into account the thermal motion and deformation of the biological medium, based on the idea of the double spring model in our previous study\cite{Lu2022Double}, we propose a soft matter potential (SMP) in this paper.

By using Brownian dynamics simulations, the temporal evolution of velocity and diffusivity is studied.
We find that the diffusivity $D(t)$ and the velocity $v(t)$ exhibit transient non-linearity at intermediate time scales.
This transient non-linearity exhibits different features as the external force $F_e$ increases.
For the velocity, as the external force increases, the particles exhibit a transition from sub-transport to free transport at intermediate time scales. For $D(t)$, as the external force increases, the particles exhibit a three-stage transition from sub-diffusion to super-diffusion to free diffusion (with weak oscillations) at intermediate time scales.
The super-diffusion phenomenon of the particles occurs when the external force is comparable to the critical force, i.e., $\tilde{F} \approx 1$, which leads to the GAD phenomenon.
Compared to WBP, the particles in the SMP exhibit a weaker GAD phenomenon, i.e., a smaller diffusivity peak, which is caused by the deformability of the biological medium. 

Based the theory of Adrover {\it et.al.} \cite{Adrover2019Exact,Adrover2019Laminar}, the normalized velocity $\tilde{v}(\tilde{F})$ and diffusivity $\tilde{D}(\tilde{F})$ of particles in SMP with different parameters is analyzed.
We investigate the effects of the stiffness $W_0$ and the thermal motion intensity $D_c$ of the biological medium on the $\tilde{v}(\tilde{F})$ and $\tilde{D}(\tilde{F})$ when the interaction $V_0$ between the particle and the biological medium is fixed.
Theoretical predictions show that decreasing $W_0$ or increasing $D_c$ can make the biological medium tend to a simple fluid.
Conversely, the SMP tend to a WBP with the same $V_0$, as $W_0$ increases or $D_c$ decreases.

The results of this paper reveal the effects of the stiffness and thermal motion intensity of the complex medium on the transport properties, which will help the design of drug delivery and drug release. 
A controllable drug release can be based on the joint regulation of the medium properties and the external physical field.
\begin{acknowledgments}
This research is supported by the Natural Science Foundation
of China (Nos. 12332016, 11832017 and 12172209).
\end{acknowledgments}

\section*{Data Availability}
The data that support the findings of this study are available from the corresponding author upon reasonable request.
\bibliographystyle{unsrt}
\bibliography{aipsamp}

\begin{thebibliography}{10}

\bibitem{Natalie2022Massively}
Natalie Boehnke, Joelle~P. Straehla, Hannah~C. Safford, Mustafa Kocak,
  Matthew~G. Rees, Melissa Ronan, Danny Rosenberg, Charles~H. Adelmann,
  Raghu~R. Chivukula, Namita Nabar, Adam~G. Berger, Nicholas~G. Lamson,
  Jaime~H. Cheah, Hojun Li, Jennifer~A. Roth, Angela~N. Koehler, and Paula~T.
  Hammond.
\newblock Massively parallel pooled screening reveals genomic determinants of
  nanoparticle delivery.
\newblock {\em Science}, 377(6604):eabm5551, 2022.

\bibitem{Kim2022Permeability}
Won~Kyu Kim, Sebastian Milster, Rafael Roa, Matej Kandu\u{c}, and Joachim
  Dzubiella.
\newblock Permeability of polymer membranes beyond linear response.
\newblock {\em Macromolecules}, 55(16):7327--7339, 2022.

\bibitem{wang2013bursts}
Bo~Wang, James Kuo, and Steve Granick.
\newblock Bursts of active transport in living cells.
\newblock {\em Physical Review Letters}, 111(20):208102, 2013.

\bibitem{Hamann2019Nucleic}
Andrew Hamann, Albert Nguyen, and Angela Pannier.
\newblock Nucleic acid delivery to mesenchymal stem cells: a review of nonviral
  methods and applications.
\newblock {\em Journal of Biological Engineering}, 13(7), 01 2019.

\bibitem{Ellis2001Macromolecular}
R.John Ellis.
\newblock Macromolecular crowding: an important but neglected aspect of the
  intracellular environment.
\newblock {\em Current Opinion in Structural Biology}, 11(1):114--119, 2001.

\bibitem{Nakano2014EffectsOM}
Shuichi Nakano, Daisuke Miyoshi, and Naoki Sugimoto.
\newblock Effects of molecular crowding on the structures, interactions, and
  functions of nucleic acids.
\newblock {\em Chemical Reviews}, 114 5:2733--58, 2014.

\bibitem{Tan2013MolecularCS}
Cheemeng Tan, Saumya Saurabh, Marcel~P. Bruchez, Russell Schwartz, and Philip
  LeDuc.
\newblock Molecular crowding shapes gene expression in synthetic cellular
  nanosystems.
\newblock {\em Nature Nanotechnology}, 8:602 -- 608, 2013.

\bibitem{Nakano2017Model}
Shuichi Nakano and Naoki Sugimoto.
\newblock Model studies of the effects of intracellular crowding on nucleic
  acid interactions.
\newblock {\em Molecular Biosystems}, 13:32--41, 2017.

\bibitem{Si2022Modulation}
Dongqing Si, Xinyue Liu, Jinbo Wu, and Guohui Hu.
\newblock Modulation of {DNA} conformation in electrolytic nanodroplets.
\newblock {\em Phys. Chem. Chem. Phys.}, 24:6002--6010, 2022.

\bibitem{Si2018DNA}
Dongqing Si, Zhen Xu, Nan Nan, and Guohui Hu.
\newblock {DNA} confined in a nanodroplet: a molecular dynamics study.
\newblock {\em The Journal of Physical Chemistry B}, 122(38):8812--8818, 2018.
\newblock PMID: 30180585.

\bibitem{liu2020Jaggregate}
Yun Liu, Guangze Yang, Song Jin, Run Zhang, Peng Chen, Tengjisi, Lianzhou Wang,
  Dong Chen, David~A Weitz, and Chun-Xia Zhao.
\newblock J-aggregate-based fret monitoring of drug release from polymer
  nanoparticles with high drug loading.
\newblock {\em Angewandte Chemie}, 132(45):20240--20249, 2020.

\bibitem{Zelepukin2022FlashDR}
I.~Zelepukin, Olga~Yu. Griaznova, Konstantin~G. Shevchenko, Andrey~V. Ivanov,
  E.~Baidyuk, Natalia~B. Serejnikova, A.~B. Volovetskiy, S.~Deyev, and
  A.~Zvyagin.
\newblock Flash drug release from nanoparticles accumulated in the targeted
  blood vessels facilitates the tumour treatment.
\newblock {\em Nature Communications}, 13, 2022.

\bibitem{Culebras2021Wood}
Mario Culebras, Anthony Barrett, Mahboubeh Pishnamazi, Gavin~Michael Walker,
  and Maurice~N. Collins.
\newblock Wood-derived hydrogels as a platform for drug-release systems.
\newblock {\em ACS Sustainable Chemistry \& Engineering}, 9(6):2515--2522,
  2021.

\bibitem{Freitag2002randomized_extendedrelease}
F.~G. Freitag, S.~D. Collins, H.~A. Carlson, J.~Goldstein, J.~Saper,
  S.~Silberstein, N.~Mathew, P.~K. Winner, R.~Deaton, and K.~Sommerville.
\newblock A randomized trial of divalproex sodium extended-release tablets in
  migraine prophylaxis.
\newblock {\em Neurology}, 58(11):1652--1659, 2002.

\bibitem{KANE2007Treatment}
J.~Kane, F.~Canas, M.~Kramer, L.~Ford, C.~Gassmann-Mayer, P.~Lim, and
  M.~Eerdekens.
\newblock Treatment of schizophrenia with paliperidone extended-release
  tablets: A 6-week placebo-controlled trial.
\newblock {\em Schizophrenia Research}, 90(1):147--161, 2007.

\bibitem{Dell2013Theory}
Zachary~E. Dell and Kenneth~S. Schweizer.
\newblock Theory of localization and activated hopping of nanoparticles in
  cross-linked networks and entangled polymer melts.
\newblock {\em Macromolecules}, 47(1):405--414, 2013.

\bibitem{Xue2020Diffusion}
Chundong Xue, Xinghua Shi, Yu~Tian, Xu~Zheng, and Guoqing Hu.
\newblock Diffusion of nanoparticles with activated hopping in crowded polymer
  solutions.
\newblock {\em Nano Letters}, 20(5):3895--3904, 2020.

\bibitem{cai2015hopping}
Li-Heng Cai, Sergey Panyukov, and Michael Rubinstein.
\newblock Hopping diffusion of nanoparticles in polymer matrices.
\newblock {\em Macromolecules}, 48(3):847--862, 2015.

\bibitem{Lu2021potential}
Yu~Lu and Guo-Hui Hu.
\newblock A potential barrier in the diffusion of nanoparticles in ordered
  polymer networks.
\newblock {\em Soft Matter}, 17:6374--6382, 2021.

\bibitem{Kramers1940Brownian}
H.A. Kramers.
\newblock Brownian motion in a field of force and the diffusion model of
  chemical reactions.
\newblock {\em Physica}, 7(4):284--304, 1940.

\bibitem{Gang1996Diffusion}
Hu~Gang, A.~Daffertshofer, and H.~Haken.
\newblock Diffusion of periodically forced {B}rownian particles moving in
  space-periodic potentials.
\newblock {\em Physical Review Letters}, 76:4874--4877, Jun 1996.

\bibitem{Reimann2002Diffusion}
P.~Reimann, C.~Van~den Broeck, H.~Linke, P.~H\"anggi, J.~M. Rubi, and
  A.~P\'erez-Madrid.
\newblock Diffusion in tilted periodic potentials: Enhancement, universality,
  and scaling.
\newblock {\em Phys. Rev. E}, 65:031104, Feb 2002.

\bibitem{Lindenberg2005Transport}
Katja Lindenberg, A~M Lacasta, J~M Sancho, and A~H Romero.
\newblock Transport and diffusion on crystalline surfaces under external
  forces.
\newblock {\em New Journal of Physics}, 7(1):29, jan 2005.

\bibitem{Lindenberg2007Dispersionless}
Katja Lindenberg, J.~M. Sancho, A.~M. Lacasta, and I.~M. Sokolov.
\newblock Dispersionless transport in a washboard potential.
\newblock {\em Phys. Rev. Lett.}, 98:020602, Jan 2007.

\bibitem{Reimann2008Weak}
Peter Reimann and Ralf Eichhorn.
\newblock Weak disorder strongly improves the selective enhancement of
  diffusion in a tilted periodic potential.
\newblock {\em Physical Review Letters}, 101:180601, Oct 2008.

\bibitem{Burada2009Diffusion}
P.~Sekhar Burada, Peter H\"anggi, Fabio Marchesoni, Gerhard Schmid, and Peter
  Talkner.
\newblock Diffusion in confined geometries.
\newblock {\em ChemPhysChem}, 10(1):45--54, 2009.

\bibitem{Xiao2019Investigation}
Xiao-Yang Shi and Jing-Dong Bao.
\newblock Investigation on the enhancement phenomenon of biased-diffusion in
  periodic potential.
\newblock {\em Physica A: Statistical Mechanics and its Applications},
  514:203--210, 2019.

\bibitem{hanggi1990reaction}
Peter H{\"a}nggi, Peter Talkner, and Michal Borkovec.
\newblock {Reaction-rate theory: fifty years after Kramers}.
\newblock {\em Reviews of Modern Physics}, 62(2):251, 1990.

\bibitem{Reimann2001Giant}
P.~Reimann, C.~Van~den Broeck, H.~Linke, P.~H\"anggi, J.~M. Rubi, and
  A.~P\'erez-Madrid.
\newblock Giant acceleration of free diffusion by use of tilted periodic
  potentials.
\newblock {\em Physical Review Letters}, 87:010602, Jun 2001.

\bibitem{Bellando2022Giant}
L.~Bellando, M.~Kleine, Y.~Amarouchene, M.~Perrin, and Y.~Louyer.
\newblock Giant diffusion of nanomechanical rotors in a tilted washboard
  potential.
\newblock {\em Phys. Rev. Lett.}, 129:023602, Jul 2022.

\bibitem{Barato2015Thermodynamic}
Andre~C. Barato and Udo Seifert.
\newblock Thermodynamic uncertainty relation for biomolecular processes.
\newblock {\em Physical Review Letters}, 114:158101, Apr 2015.

\bibitem{horowitz2020thermodynamic}
Jordan~M Horowitz and Todd~R Gingrich.
\newblock Thermodynamic uncertainty relations constrain non-equilibrium
  fluctuations.
\newblock {\em Nature Physics}, 16(1):15--20, 2020.

\bibitem{Koyuk2020Thermodynamic}
Timur Koyuk and Udo Seifert.
\newblock Thermodynamic uncertainty relation for time-dependent driving.
\newblock {\em Phys. Rev. Lett.}, 125:260604, Dec 2020.

\bibitem{Hyeon2017Physical}
Changbong Hyeon and Wonseok Hwang.
\newblock Physical insight into the thermodynamic uncertainty relation using
  {B}rownian motion in tilted periodic potentials.
\newblock {\em Physical Review E}, 96:012156, Jul 2017.

\bibitem{Frenken1985Observation}
Joost W.~M. Frenken and J.~F. van~der Veen.
\newblock Observation of surface melting.
\newblock {\em Physical Review Letters}, 54:134--137, Jan 1985.

\bibitem{Lee2006Giant}
Sang-Hyuk Lee and David~G. Grier.
\newblock Giant colloidal diffusivity on corrugated optical vortices.
\newblock {\em Physical Review Letters}, 96:190601, May 2006.

\bibitem{Xu2021Enhanced}
Ziyang Xu, Xiaobin Dai, Xiangyu Bu, Ye~Yang, Xuanyu Zhang, Xingkun Man, Xinghua
  Zhang, Masao Doi, and Li-Tang Yan.
\newblock Enhanced heterogeneous diffusion of nanoparticles in semiflexible
  networks.
\newblock {\em ACS Nano}, 15(3):4608--4616, 2021.
\newblock PMID: 33625839.

\bibitem{Lu2022Double}
Yu~Lu, Xin-Yue Liu, and Guo-Hui Hu.
\newblock Double-spring model for nanoparticle diffusion in a polymer network.
\newblock {\em Macromolecules}, 55(11):4548--4556, 2022.

\bibitem{Adrover2019Exact}
Alessandra Adrover, Chiara Passaretti, Claudia Venditti, and Massimiliano
  Giona.
\newblock {Exact moment analysis of transient dispersion properties in periodic
  media}.
\newblock {\em Physics of Fluids}, 31(11), 2019.

\bibitem{Adrover2019Laminar}
Alessandra Adrover, Claudia Venditti, and Massimiliano Giona.
\newblock {Laminar dispersion at low and high Peclet numbers in a sinusoidal
  microtube: point-size versus finite-size particles}.
\newblock {\em Physics of Fluids}, 31(6), 2019.

\bibitem{wang2022double}
Cun-Hai Wang, Zi-Yang Liu, Ze-Yi Jiang, and Xin-Xin Zhang.
\newblock Double-diffusive convection in a magnetic nanofluid-filled porous
  medium: Development and application of a nonorthogonal lattice boltzmann
  model.
\newblock {\em Physics of Fluids}, 34(6), 2022.

\bibitem{Jakhar2023Instability}
Atul Jakhar and Anand Kumar.
\newblock {Instability analysis of double diffusive convection under time
  dependent solute boundary conditions in the presence of internal heat
  generator}.
\newblock {\em Physics of Fluids}, 35(7):077101, 07 2023.

\bibitem{Vargas2023ColloidPOF}
C.~Vargas, F.~M\'endez, A.~Docoslis, and C.~Escobedo.
\newblock {Colloid transport by an oscillatory electroosmotic flow between
  microelectrodes of axially variable shape}.
\newblock {\em Physics of Fluids}, 35(9):092014, 09 2023.

\bibitem{Venditti2022Exactdispersion}
Claudia Venditti, Massimiliano Giona, and Alessandra Adrover.
\newblock {Exact moment analysis of transient/asymptotic dispersion properties
  in periodic media with adsorbing/desorbing walls}.
\newblock {\em Physics of Fluids}, 34(12):122013, 12 2022.

\bibitem{Venditti2022ComparisonPOF}
Claudia Venditti, Stefano Cerbelli, Giuseppe Procopio, and Alessandra Adrover.
\newblock {Comparison between one- and two-way coupling approaches for
  estimating effective transport properties of suspended particles undergoing
  Brownian sieving hydrodynamic chromatography}.
\newblock {\em Physics of Fluids}, 34(4):042010, 04 2022.

\bibitem{Venditti2022Physica}
Claudia Venditti, Alessandra Adrover, and Massimiliano Giona.
\newblock {Inertial effects and long-term transport properties of particle
  motion in washboard potential}.
\newblock {\em Physica A: Statistical Mechanics and its Applications},
  585:126407, 2022.

\bibitem{Brenner1980Dispersion}
H.~Brenner and Keith Stewartson.
\newblock Dispersion resulting from flow through spatially periodic porous
  media.
\newblock {\em Philosophical Transactions of the Royal Society of London.
  Series A, Mathematical and Physical Sciences}, 297(1430):81--133, 1980.

\bibitem{hofling2013anomalous}
Felix H{\"o}fling and Thomas Franosch.
\newblock Anomalous transport in the crowded world of biological cells.
\newblock {\em Reports on Progress in Physics}, 76(4):046602, 2013.

\end{thebibliography}

\end{document}